\documentclass[conference]{IEEEtran}
\IEEEoverridecommandlockouts
\usepackage{cite}
\usepackage{amsmath,amssymb,amsfonts}
\usepackage{algorithmic}
\usepackage[ruled]{algorithm2e}
\usepackage{graphicx}
\usepackage{subfigure}
\usepackage{epstopdf}
\usepackage{textcomp}
\usepackage{xcolor}
\usepackage{multirow}
\def\BibTeX{{\rm B\kern-.05em{\sc i\kern-.025em b}\kern-.08em
    T\kern-.1667em\lower.7ex\hbox{E}\kern-.125emX}}
\begin{document}

\title {A More Refined Mobile Edge Cache Replacement Scheme for Adaptive Video Streaming with Mutual Cooperation in Multi-MEC Servers }

\author{\IEEEauthorblockN{ Xinyu Huang, Lijun He$^{*}$, Xin Chen, Fan Li, Guizhong Liu}
\IEEEauthorblockA{\textit{The Department of Electronic and Information Engineering, Xi'an Jiao Tong University, Xi'an, China} \\
\{xinyu\_huang, xing\_chen\}@stu.xjtu.edu.cn, \{jzb2016125, lifan, liugz\}@mail.xjtu.edu.cn }}

\maketitle

\begin{abstract}
In this paper, we propose a more refined video segment based Mobile Edge Computing (MEC) enhanced cache update strategy, which takes into account the client's playback status and transmission state, MEC cache capacity and the popularity of each segment, to improve the quality of experience (QoE) of clients and the use ratio of MEC cache. In each cache update period, the segments which cannot bring significant QoE improvement will be deleted and meanwhile more suitable representations of the segments will be cahed instead. First, we divide the MEC cache space into three parts based on segment popularity and segment importance level. In addition, the size of different cache parts can be transformed to each other to further exploit the MEC cache space. For different parts, the corresponding deleted strategy and caching strategy are formulated, where the deleted strategy combines the requested times of segment and transmission capability of clients, and the caching strategy utilizes the playback state of clients to calculate the caching priority. Furthermore, we formulate the cache update problem to maximize the utility function of clients subject to the constraint of MEC cache size and transmission capacity. The brand and branch method is emplyed to obatain the optimal solution. Simulation results show that our proposed algorithm can improve system throughput and hit ratio of video segment, and at the same time decrease playback frozen time and system backhaul traffic compared with other existing algorithms.
\end{abstract}


\section{Introduction}
As the key technology of the fifth generation mobile communication system \cite{b1}, MEC server is closer to the clients' nodes so that they can obtain client information (e.g. client's requirements, network state and playback status) in real time. In addition, MEC server owns powerful storage and computation capacity. Pre-caching some video contents at MEC server can avoid the repeated video content transmission and decrease the burden on the backhaul network, thereby achieving faster service response and improving client's QoE \cite{b2}. In this paper, our work is to design a more refined segment-based cache update strategy, to improve the system throughput, the hit ratio of the segments, as well as to reduce playback frozen time and the backhaul traffic.

To further improve the cache efficiency of the MEC, the researchers proposed many video cache update algorithms. In \cite{b3}, the authors introduced the methods to predict video popularity distribution and client preferences, and proposed a wireless edge caching and video content transmission framework. However, cooperative caching schemes among multiple MEC servers has not been considered. In \cite{b4}, the best video representation was cached for each MEC server to minimize the aggregate average video distortion reduction of all clients. Nevertheless, the impact of video representation switch on client's QoE has not been considered. The authors in \cite{b5} proposed a MEC cache update algorithm based on the video popularity and wireless channel state, to maximize the use ratio of the MEC cache and reource efficiency. However, it did not consider the client's playback status, which can result in that frequent representation switch was negelected.
The authors in \cite{b6} proposed an network-aware edge cache update algorithm, which integrated client's channel state and segment popularity. But the cache optimization and the clients' plaback status was not jointly considered, which may cause the rebuffering event. In \cite{b7}, the authors selected video rate, initial delay, interruption time, representation switch and fairness as the cache update indicator, and explored the trade-off between QoE and backhaul traffic. In \cite{b8}, the authors adopted the method of caching the most popular video that can be supported according to the throughput of the current network, and guaranteed maximum video playback quality and minimal quality fluctuation. However, the disadvantage of them is that the deleted policy only made decision based on the number of segment requests in the current update period without considering the overall popularity, which may cause additional burden on the backhaul network. 

In general, there are still some problems to be solved for the MEC video cache update strategy: (1) The importance of the first few segments of video has not been considered, which results in a large initial delay. (2) The client's playback status and channel state have been neglected, which leads to frequent rebuffering events. (3) Collaboration among MEC servers has not been exploited and thus the video information is not shared. Inspired by the above work, we propose a more refined MEC cache update strategy to improve system throughput and hit ratio of segment, as well as reduce backhaul traffic, client's playback frozen time. Our work is novel in the following aspects.
\begin{itemize}
 \item 	Refined MEC cache partition based on content characteristic and segment popularity
 
According to the characteristic of each segment, the MEC cache is divided into a high-popular and high-important one, a high-popular but low-important one and a low-popular one, respectively, to avoid additional backhaul traffic when frequent updates.

\item The calculation of the segment delete priority by combining segment request number and client's transmission state

To avoid the situation that the cached segment cannot match the client's transmission capability, we combine the number of segment requests in the current update period with the client's transmission capability, to calculate the delete priority of the segment as the basis to determine which segment should be deleted.

\item Cache utility function and optimal solution

By comprehensively considering the client's playback status, channel state and cooperation mechanism among MEC servers, we obtain the cache utility function, which is employed to formulate the objective problem. To reduce the computational complexity, we transform the client's cache problem into segment cache problem and utilize the brand and branch method to achieve the optimal solution.

\end{itemize}

The rest of the paper is organized as follows: Section II introduces the proposed system model. Section III elaborates the MEC cache update strategy from aspects of the MEC cache area partition, segment update strategy, MEC cache space transfer and MEC collaboration mechanism respectively. Section IV formulates the cache update problem and obtain the optimal solution. The simulation results and analysis are given in Section V. Finally, we draw the conclusion of the paper in Section VI.

\section{System Description}

\subsection{System Framework}

\begin{figure}[!h]
	\centering
	\includegraphics[width=0.46\textwidth]{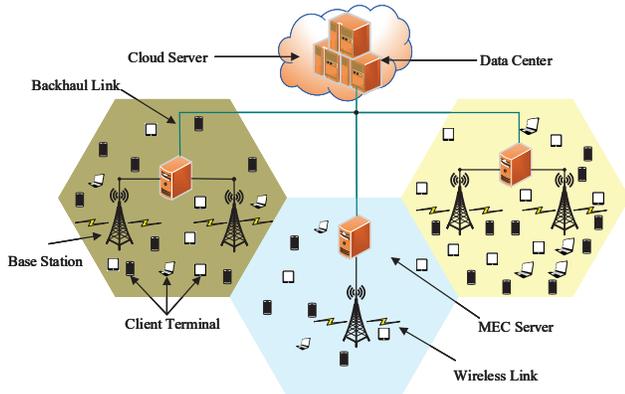}
	\caption{System Model}
	\label{fig:system}
\end{figure}

In the scenario of this paper, when multiple wireless client terminals request video services, the local MEC server will provide video cache services to the neighboring eNodeBs, and the cooperative MEC servers will share video stream with each other, as shown in Fig. \ref{fig:system}.

The MEC system is mainly composed of cloud server, MEC server, client terminals and eNodeB. The cloud server is the data center to provide all the videos to clients. The MEC server has powerful storage and computing capacity, which can be used to cache and transcode videos, respectively. First, the eNodeB sends a video request to the local MEC server according to the request of the client. If the video stream has been cached in the local MEC server, it will be directly transmitted to the eNodeB and then delivered to the requested client. Otherwise, the local MEC server will send the video request information to the cooperative MEC servers for sharing video stream or request the video stream from the cloud server. 

\subsection{The Setting of MEC Cache Size}

In the multi-MEC servers scene, we denote the set of MEC servers by $ \mathbb {Q}=\{1,\cdots,q, \cdots, Q\}$  and divide the scene into different regions based on coverage. Each region $\mathfrak{R}_q$ with $K_q$ clients is served by one MEC server and several eNodeBs. To be specific, the corresponding eNodeB set of MEC q is $\mathbb{H}_q=\{1,\cdots,h_q,\cdots, H_q\}$, and its served client set is  $\mathbb{Z}(h_q)$, which satisfies  $\sum_{h_q=1}^{H_q}\mathbb{Z}(h_q)=K_q$. Multiple eNodeBs connect to the MEC server to request video through wired link. To determine the setting of MEC server cache size, historical analysis is performed according to the amount of video provided by the eNodeBs in each region, to meet the needs of clients in different regions.

It is assumed that the total cache size of MEC servers is $S$. Then the cache size $S_q$ of different MEC servers is set according to the statistical analysis of the video in the recent month, which can be written as:
\begin{equation}
S_q=\frac{\sum_{k\in\mathfrak{R}_q}\sum_{f\in\mathcal{F}}\sum_{i\in SR_{k,f}}\Gamma_{k,f,i}^{l}s_{f,i}^{l}}{\sum_{k\in\mathfrak{R}}\sum_{f\in\mathcal{F}}\sum_{i\in SR_{k,f}}\Gamma_{k,f,i}^{l}s_{f,i}^{l}}S
\end{equation}
where $\mathcal{F}=\{1,\cdots,f,\cdots, F\}$ denotes the set of video files.

$\Gamma_{k,f,i}^{l}$ indicates the request number of client k for the representation $l$ of the segment $i$ of the video $f$ within one month. $s_{f,i}^{l}$ and $SR_{k,f}$ denote the size of the representation $l$ of the segment $i$ of the video $f$ and the set of requested segments of video $f$ respectively.

\section{MEC Cache update Strategy}
In each cache update period, the delete priority of each representation of the segments in the local MEC server is calculated to perform the deletion policy. The cache policy is based on the cache priority of the segment requested by the client and collaboration among MEC servers, of which the purpose is to cache the required segments with higher cache priority and delete the ones with high delete priority in each cache scheduling period in the case of not exceeding the transmission capacity.

\subsection{MEC Cache Partition}
In the initial stage, since the number of videos clients request is large and the cache space of the MEC server is limited, the first $e_f$ segments with the highest popularity representation of the video $f$ are placed into the MEC cache space based on the descending popularity of the videos until the cache of MEC server is full, in order to reduce the initial delay and improve the QoE of clients. The statistical result shows that the stop watching probability of the first $15\%$ of the video content is 0.5, so the video content of the first $15\%$ of the video is more important\cite{b10}. Therefore, we can have $e_f=\left \lceil 0.15N_f \right \rceil $, where $N_f$ is the number of segments of video $f$.

\begin{figure}[!h]
	\centering
	\includegraphics[width=0.5\textwidth]{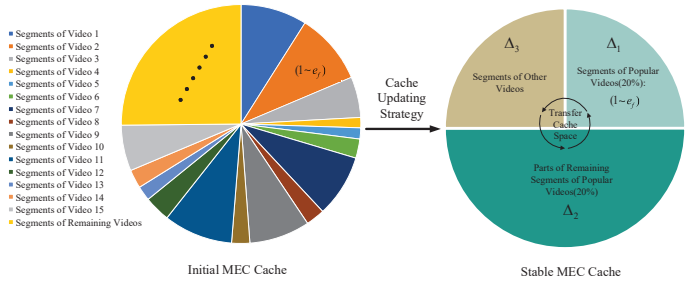}
	\caption{MEC Cache Partition}
	\label{fig:cache_state}
\end{figure}

If all stored segments are updated in real time, the segments being deleted in the current period may need to be cached in the next period, which increases the backhaul traffic pressure. Therefore, we first set different lengths of update period according to the characteristics of the segment content. And then we divide the MEC cache into three categories and utilize $\Delta_1,\Delta_2,\Delta_3$ to represent: (1) the first $15\%$ segments of the popular videos constitute the first part of the cache $\Delta_1$, (2) the remaining segments of the popular videos constitute $\Delta_2$, (3) the segments of non-popular videos compose $\Delta_3$. The cache $\Delta_1$ is updated every long period $\mathcal{J} \in\{1,2,\cdots,\Theta\}$, and the cache $\Delta_2$ and $\Delta_3$ are updated every short period $\gamma\in\{1,2,\cdots,\Upsilon\}$, as shown in Fig.\ref{fig:cache_state}.

\subsection{Segment update Strategy}

First, we employ the request number, client's transmission capacity and segment rate to obtain the delete priority of the different representations of the segments in the period $\gamma$ for the MEC server $q$, which can be depicted as
\begin{equation}
DP_i^{\gamma,q,l}=\frac{1}{K_q}\sum_{k=1}^{K_q}\frac{1}{g_{k,i}^{\gamma,l}+\zeta\cdot v_{k,i}^{l}+\alpha}
\end{equation}
where $\zeta$ and $\alpha$ are the positive variables. $v_{k,i}^{l}$ reflects the match relationship between average transmission capacity, $\widehat{CP}_{k,i}^{l}$, and segment rate, $R_{i}^{l}$, when the client $k$ receives the segment $i$ of representation $l$.
\begin{equation}
v_{k,i}^{l}=\left\{\begin{matrix}
1,&\widehat{CP}_{k,i}^{l}\geq R_{i}^{l}\\ 
0,&otherwise
\end{matrix}\right.
\end{equation}

However, it's inaccurate to perform the delete strategy only based on the delete priority in the current period $\gamma$, which may delete the segment of low delete priority in the previous periods. Therefore, we employ the delete strategy of pervious period to update the current period, which can be calculated by
\begin{equation}\label{eq:DP}
\widehat{DP}_{i}^{\gamma,q,l}=\lambda DP_{i}^{\gamma,q,l}+(1-\lambda) \widehat{DP}_{i}^{\gamma -1,q,l}
\end{equation} 

The segment update strategy is various in parts of cache. For cache $\Delta_1$, since the first $e_f$ segments are more important to clients, we will update them according to the latest video popularity each period $\mathcal{J}$. For cache $\Delta_2$ and $\Delta_3$, due to the limited transmission capacity of the backhaul, it's hard to cache all requested segments every cache update period. Therefore, the cache priority of clients is considered to decide which segment should be cached. In addition, if the delete and cache tasks are conducted separately in the different parts of cache, the benefit of delete and cache cannot be maximized. Therefore, we jointly optimize the cache update strategy in cache $\Delta_2$ and $\Delta_3$. 

Cache strategy: Since the clients' remaining buffer are generally small, it is better to distinguish each client's urgency according to corresponding playback buffer status. Segmentation exponential function is used to indicate the cache priority of client $k$ requesting segment $i$ of representation $l$, which can be expressed as
\begin{equation}
Pr_{k,i}^{\gamma,l}=\left\{\begin{matrix}
0, & BT_{k,i-1}^{\gamma,l}\geq \frac{s_{k,i}^{l}}{\widehat{CP}_{k,i-1}^{l}}\\ 
e^{\frac{\omega}{1+BT_{k,i-1}^{\gamma,l}}}-1, & otherwise 
\end{matrix}\right.
\end{equation}

$BT_{k,i-1}^{\gamma,l}$ is the remaining buffer time when client $k$ receives the segment $i-1$ of representation $l$ at the end of period $\gamma$, which can be calculated by $BT_{k,i-1}^{\gamma,l}=\mu_k /FR$. $\mu_k$ is the number of unplayed frames of all completely received segments in client $k$'s buffer, and $FR$ is the playback frame rate. If the remaining playback time of the client exceeds the transmission time of the requested segment, we classify the client $k$ as a non-urgency client and the client priority is set to $0$.

Delete strategy: In each cache update period $\gamma$, the stored segments in $\Delta_2$ and $\Delta_3$ are sequentially deleted based on the overall delete priority of segments excluding the segments being transmitted, until the size of cache area is large enough to store the segments to be downloaded.

\subsection{MEC Cache Space Transfer}
The segments in $\Delta_1$ are updated every period $\mathcal{J}$. Due to the length and code rate, the size of segments differs a lot, which will results in that the size of delete segments cannot equal the downloaded segments. To make full use of each cache space, we compare the size of $\Delta_1$ before and after each update period, to release or increase some space of $\Delta_2$ and $\Delta_3$. In the beginning of each period $\mathcal{J}$, the size of $\Delta_1$ is derived by
\begin{equation}
FC_{q}^{\mathcal{J}}=\sum_{f\in \mathcal{F}_{pop}^{\mathcal{J}}}\sum_{i=1}^{e_f}\sum_{l=1}^{L}s_{f,i}^{l}-SH_{q}^{\mathcal{J}-1}
\end{equation}
where $\mathcal{F}_{pop}^{\mathcal{J}}$ denotes the set of popular videos. $SH_{q}^{\mathcal{J}-1}$ denotes the size of $\Delta_1$ in the period $\mathcal{J}-1$. Then, the sum size of $\Delta_2$ and $\Delta_3$ can be updated according to $FC_{q}^{\mathcal{J}}$ as
\begin{equation}
SC_{q}^{\mathcal{J}}=SC_{q}^{\mathcal{J}-1}-FC_{q}^{\mathcal{J}}
\end{equation}
When $FC_{q}^{\mathcal{J}}\geq 0$, the segments in $\Delta_2$ and $\Delta_3$ will be deleted according to the delete priority to release the cache to $\Delta_1$ until the size of $\Delta_1$ is satisfied. Acnd $FC_{q}^{\mathcal{J}}<0$ otherwise. 

\subsection{Collaboration mechanism among MEC Servers}

If the requested segments have not been stored in the local MEC server $q$, the requested segment information will be first sent to the neighboring MEC servers. Provided that the neighboring MEC servers store the requested segments, the local MEC server will download them from the neighboring MEC server $p$ with the highest remaining transmission capacity. The sum size of downloaded segments should not exceed the transmission capacity between the local MEC server $q$ and the neighboring MEC server $p$.

\begin{equation}
\sum_{k=1}^{K_q}\sum_{l=1}^{L}\tau_{k,i,l}^{q}\chi_{k,i,l}^{p}s_{i}^{l}\leq CP_{q,p}\cdot TD, \forall q\in \mathbb{Q}, p\in \mathbb{Q}\setminus\{q\}
\end{equation}
where $\tau_{k,i,l}^{q}$ denotes the optimized variable, which indicates whether the requested segment of client $k$ will be cached in MEC server $q$. $\tau_{k,i,l}^{q}=1$ represents that the segment $i$ of representation $l$ should be cached in the MEC server $q$ and $\tau_{k,i,l}^{q}=0$ otherwise. $\chi_{k,i,l}^{q}$ denotes whether the segment $i$ of representation $l$ is stored in MEC server $p$. If it is stored, $\chi_{k,i,l}^{p}=1$, otherwise, $\chi_{k,i,l}^{p}=0$. $CP_{q,p}$ is the transmission capacity between MEC server $q$ and $p$. $TD$ is the time duration of period $\gamma$. If the transmitted segments exceed the transmission capacity, the segment share mechanism among MEC servers will be stopped and the local MEC server $q$ should request the segment from the cloud server or wait until the next update period.

\section{Problem Formulation and Solution Method}

\subsection{Utility Function of Client Cache Priority}
In each cache update period, MEC server needs to cache the suitable representation of segments for the clients with high cache priority. Since the time of segments from the cloud server and the neighboring MEC server is different, when the requested segment is downloaded from the cloud server, the cache priority of client $k$ needs to be improved, otherwise remains unchanged. Then, the utility function of client's cache priority can be defined as 
\begin{equation}
U_{k,i,l}=Pr_{k,i,l}+Pr_{k,i,l}\prod_{p}(1-\chi_{k,i,l}^{p})
\end{equation}

\subsection{Problem Formulation}
In the cache $\Delta_1$, the cache update strategy is that the first $e_f$ segments of popular videos are updated every period $\mathcal{J}$, which is relatively fixed. In the cache $\Delta_2$ and $\Delta_3$, the goal is to cache as many segments requested by clients of high cache priority as possible, which can be conversed to maximize the total cache utility of clients as follows: 
\begin{equation}\scriptsize\label{eq:program}
\begin{split}
& \max\sum_{k\in \mathfrak{R}_q}\sum_{l=1}^{L}\tau_{k,i,l}^{q}\cdot U_{k,i,l}\\ 
&s.t. \\ 
&(c1)\sum_{k\in \mathfrak{R}_q}\sum_{l=1}^{L}\tau_{k,i,l}^{q}s_{i}^{l}\leq SC_q , \forall q\in \mathbb{Q}\\ 
&(c2)\tau_{k,i,l}^{q}+\chi_{k,i,l}^{q}\leq 1 , \forall q\in \mathbb{Q}\\ 
&(c3)\sum_{k\in \mathfrak{R}_q}\sum_{l=1}^{L}\tau_{k,i,l}^{q}\chi_{k,i,l}^{p}s_{i}^{l}\leq CP_{q,p} TD , \forall q\in \mathbb{Q},p\in \mathbb{Q}\setminus\{q\}\\ 
&(c4)\sum_{k\in \mathfrak{R}_q}\sum_{l=1}^{L}\tau_{k,i,l}^{q}\prod_{p}(1-\chi_{k,i,l}^{p})s_{i}^{l}\leq CP_{q,ori} TD , \forall q\in \mathbb{Q},p\in \mathbb{Q}\setminus\{q\}
\end{split}
\end{equation}
Constraint (c1) implies that the segments to be cached should not be greater than the size of $\Delta_2$ and $\Delta_3$. Constraint (c2) guarantees that the segments cannot be cached repeatedly. Constraint (c3) and constraint (c4) indicate that the size of segments to be downloaded from the neighboring MEC server and the cloud server should not exceed the corresponding transmission capacity respectively.

\subsection{Solution Method}
The optimized problem is a 0-1 inter programming problem for the variable $\tau$. To reduce the computational complexity, the Eq.(\ref{eq:program}) can be simplified as follows.

\begin{figure}[!h]
	\centering
	\includegraphics[width=0.5\textwidth]{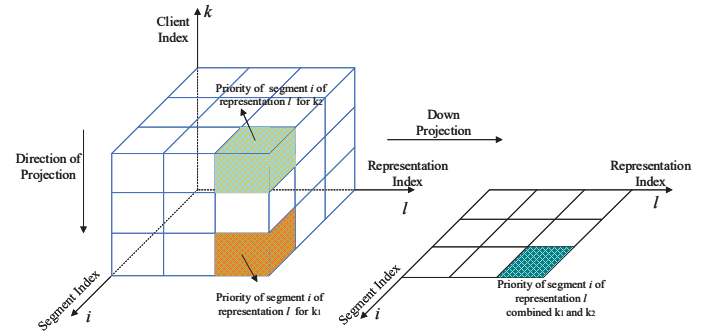}
	\caption{Narrow the Search Range}
	\label{fig:narrow}
\end{figure}

First, we can utilize the constraint (c2) to narrow the search range. The client set $E_q$ satisfying the condition ($\chi_{k,i,l}^{q}=1, \forall k\in \mathfrak{R}_q$) can be removed directly from the client set $\mathfrak{R}_q $. In the cache decision determination, the condition of one segment to be requested by many clients often happens. To further narrow the search range, we transform the client's cache problem into the segment cache problem as described in Fig.\ref{fig:narrow}.

We transform the requested segments of client set $\mathfrak{R}_q - E_q$ into a two-dimensional array, which is segment index and representation index respectively. When the segment is requested by several clients, the utility function is derived as follows. 
\begin{equation}
Ps_{i}^{l}=\sum_{k\in \mathfrak{R}_q - E_q}U_{k,i,l}
\end{equation}

Then the search range can be narrowed from $(\mathfrak{R}_q - E_q)\times L$ to $I\times L$, where $I$ is the number of requested segments. Further, the objective function can be simplified as
\begin{equation}
\max\sum_{i=1}^{I}\sum_{l=1}^{L}\tau_{i,l}^{q}\cdot Ps_{i,l}
\end{equation}

Since the $\chi_{k,i,l}^{p}s_{k,i}^{l}$ and $\prod_{p}(1-\chi_{k,i,l}^{p})s_{k,i}^{l}$ in the constraint (c3)(c4) are the known constant in the cache update period, we employ the constant $A_{i,l}^{p}$ and $B_{i,l}$ to substitute in to the Eq.(\ref{eq:program}), which can be simplified as 
\begin{equation}\label{eq:simple}
\begin{split}
& \max\sum_{i=1}^{I}\sum_{l=1}^{L}\tau_{i,l}^{q}\cdot Ps_{i,l}\\ 
&s.t. \\ 
&(c1)\sum_{i=1}^{I}\sum_{l=1}^{L}\tau_{i,l}^{q}s_{i}^{l}\leq SC_q , \forall q\in \mathbb{Q}\\ 
&(c2)\sum_{i=1}^{I}\sum_{l=1}^{L}\tau_{i,l}^{q}A_{i,l}^{p}\leq CP_{q,p} TD , \forall q\in \mathbb{Q},p\in \mathbb{Q}\setminus\{q\}\\ 
&(c3)\sum_{i=1}^{I}\sum_{l=1}^{L}\tau_{i,l}^{q}B_{i,l}\leq CP_{q,ori} TD , \forall q\in \mathbb{Q},p\in \mathbb{Q}\setminus\{q\}
\end{split}
\end{equation}

To solve the problem in Eq.(\ref{eq:simple}), we employ the brand and branch method \cite{b11} to obtain the optimal solution. The detailed process of our proposed algorithm is described in Algorithm \ref{cache}.
\begin{algorithm}[!h]
	\caption{The Proposed Cache update Algorithm}
	\label{cache}
		\LinesNumbered 
	\textbf{Initialize:} Size of free space $Free_q = 0, q\in\mathbb{Q}$. The size of downloaded segments $DL_q =0$. The cache variable $\tau_{i,l}^{q}=0$ and the deleted variable $DS_{i,l}^{q}=0$, for $1\leq i\leq I, 1\leq l \leq L$.  Time period $\gamma$ and $\mathcal{J}$.\\
	\KwOut{$\tau_{i,l}^{q}$, $DS_{i,l}^{q}$}
	\For{$\mathcal{J}\in\Theta$}
	{
	Update segments in $\Delta_1$ using the lateset video popularity;
	
	Calculate the size of $\Delta_1$, $FC_q^{\mathcal{J}}$, based on (6) ;
	
	Update the size of $\Delta_2$ and $\Delta_3$ based on (7);
	
		\For{$\gamma\in\mathcal{J}$}
		{
			Transform $\tau_{i,l}^{q}$ to the slack variable $0\leq\tau_{i,l}^{q}\leq 1$;
			
			Solve the Problem (13), and obtain the optimal solution $X^{*}$;
			
			\eIf{$X^{*}\neq \varnothing $}
			{
				\eIf{ all elements in $X^{*}$ satisfy $\tau_{i,l}^{q}=0, 1$}
				{
					 Output the cache variable $\tau_{i,l}^{q}$;
					
				}
				{
					Select the unsatisfied element $\tau_{i,l}^{q}$, and add $\tau_{i,l}^{q}=0$ or $\tau_{i,l}^{q}=1$ to the
					constraints;
					
					Update the optimal solution $X^{*}$;
					
				}	
			}			
			{
				\textbf{break}; // No cache strategy 
			}

			Calculate the size of segments to be deleted, $DL_q$, based on cache variable $\tau_{i,l}^{q}$;
			
			Sort the cached segments in $\Delta_2$ and $\Delta_3$ in descending delete priority based on (4);
			
			\For{ each cached segment in $\Delta_2$ and $\Delta_3$}
			{
				
				Delete the cached segment, $DS_{i,l}^{q}=1$ and Output the deleted variable $DS_{i,l}^{q}$;
			
				Update the size of free space, $Free_q = Free_q + s_i^l$;
				
				\If{the number of bytes to be freed is guaranteed, $Free_q\geq DL_q$}
				
				{
					break;			
				}  
			
			}
		}
	
	}

\end{algorithm}

\section{Numerical Results}

\subsection{Parameter Setting}
The specific simulation parameters are presented in Table \ref{tab:parameter}.
\begin{table}[]
	\caption{Simulation Parameter Configuration}\label{tab:parameter}
	\centering
	\begin{tabular}{|l|l|}
		\hline
		\multirow{2}{*}{Position of eNodeBs}                                                                & \multirow{2}{*}{\begin{tabular}[c]{@{}l@{}}{[}600, 342{]},{[}600, -342{]}, {[}0, -690{]}, \\ {[}-600, -342{]}, {[}-600 342{]}, {[}0, 690{]} (m)\end{tabular}} \\
		&                                                                                                                                                               \\ \hline
		Bandwidth                                                                                           & 20MHz                                                                                                                                                         \\ \hline
		Position of MEC servers                                                                             & {[}-600,0{]}, {[}0,0{]}, {[}600, 0{]} (m)                                                                                                                     \\ \hline
		Resource allocation algorithm                                                                       & Round Robin                                                                                                                                                   \\ \hline
		\begin{tabular}[c]{@{}l@{}}Transmission capacity between\\ MEC server and cloud server\end{tabular} & 500Mbps                                                                                                                                                       \\ \hline
		\begin{tabular}[c]{@{}l@{}}Transmission capacity among \\ MEC servers\end{tabular}                  & 200Mbps                                                                                                                                                       \\ \hline
		Pathloss                                                                                            & 20dB                                                                                                                                                          \\ \hline
		The number of clients                                                                                 & 378                                                                                                                                                           \\ \hline
		TTI                                                                                                 & 1ms                                                                                                                                                           \\ \hline
		Video segment encoder                                                                               & H.264/AVC                                                                                                                                                     \\ \hline
		\multirow{2}{*}{Video sequence}                                                                     & \begin{tabular}[c]{@{}l@{}}CIF format: Bus,Coastguard,Highway,\\ Flower, Foreman, Crew, News, Soccer\end{tabular}                                             \\ \cline{2-2} 
		& \begin{tabular}[c]{@{}l@{}}720P format: ParkingLot1,Stockholm,\\ Mobcal, In\_to\_tree, Park\_joy, Shields, \\ Ducks\_take\_off,Old\_town\_cross\end{tabular} \\ \hline
		\begin{tabular}[c]{@{}l@{}}The number of frames \\ in one segment\end{tabular}                      & 60                                                                                                                                                            \\ \hline
		Width$\times$Height                                                                                 & 352$\times$288,1280$\times$720                                                                                                                                \\ \hline
		QP                                                                                                  & 28,,29,30,31,32                                                                                                                                               \\ \hline
		Representation                                                                                      & 1,2,3,4,5                                                                                                                                                     \\ \hline
		Experimental parameters                                                                             & \begin{tabular}[c]{@{}l@{}}$\alpha=0.5$, $\beta=0.6$, $\zeta=0.8$, \\ $\lambda=0.8$, $\omega=2$, $TD=100$\end{tabular}                                        \\ \hline
	\end{tabular}
\end{table}
The algorithms compared in this section are as follows:

\textbf{Proposed:} The proposed cache update algorithm in this paper.

\textbf{LRU:} Least Recently Used\cite{b12}. The idea of this algorithm is that: the more the data was requested in the previous period, the more the data is requested in the later period. 

\textbf{LFU:} Least Recently Used\cite{b13}. The idea of this algorithm is that: if the data has been requested recently, its probability of being requested in the future is also higher. 

\textbf{WGDSF:} Weighted Greedy Dual Size Frequency\cite{b14}. This algorithm is an improvement on the Greedy Dual Size Frequency algorithm, which adds the factors of weighted frequency-based time and weighted document type.

\textbf{RBCC:} The cache replacement strategy which accounts for the possibility of collaborative content fetching\cite{b15}. Besides, it considers the values of the requested segments from the neighboring clients for caching at each edge server. 

\section{Performance Evaluation}

We validate the algorithm performance in term of system throughput, hit ratio of segment, playback frozen time and system backhaul traffic under different simulation configurations.

\begin{figure}[htbp]
	\centering
	\subfigure{
		\begin{minipage}[t]{0.5\linewidth}
			\centering
			\includegraphics[width=1.5in]{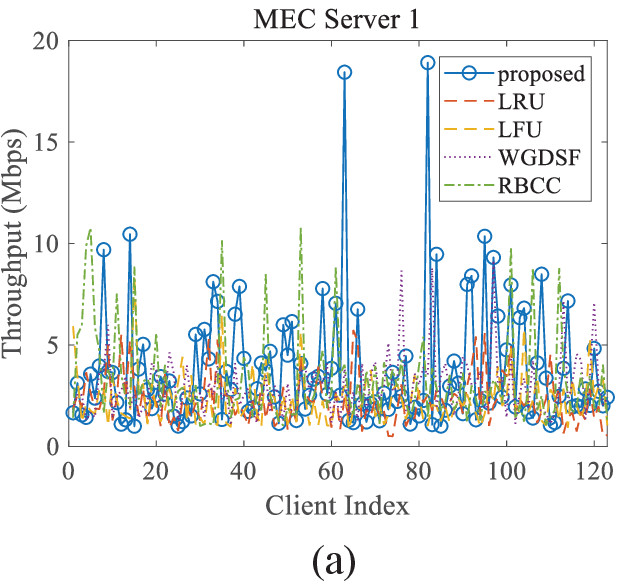}
		\end{minipage}%
	}%
	\subfigure{
		\begin{minipage}[t]{0.5\linewidth}
			\centering
			\includegraphics[width=1.5in]{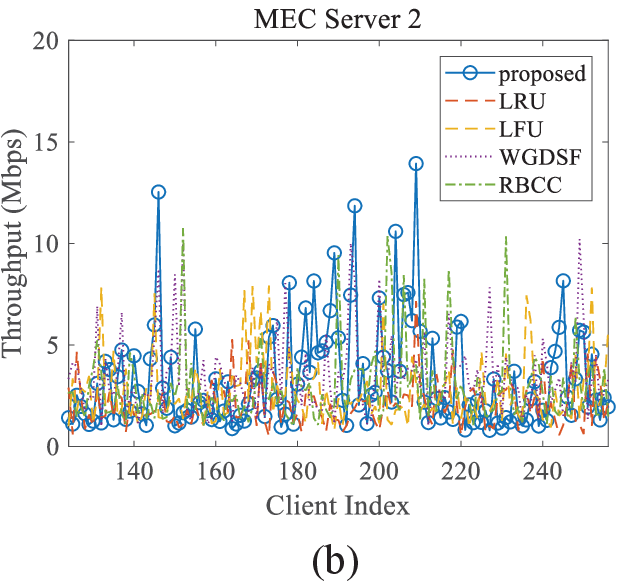}
		\end{minipage}%
	}%

	\subfigure{
		\begin{minipage}[t]{0.5\linewidth}
			\centering
			\includegraphics[width=1.5in]{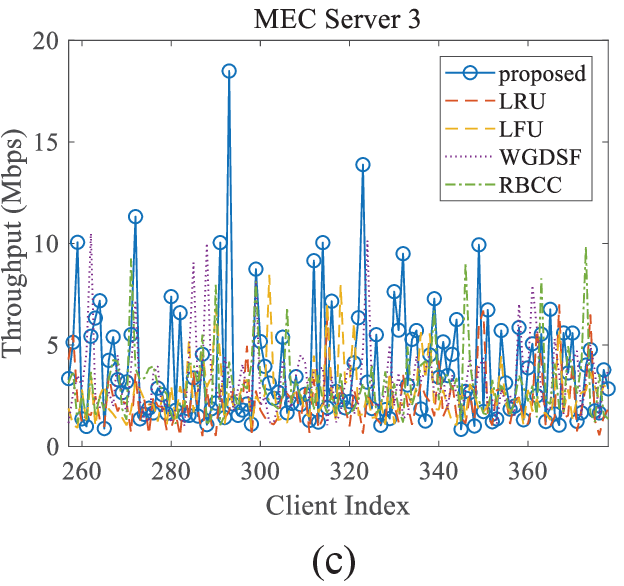}
		\end{minipage}
	}%
	\subfigure{
		\begin{minipage}[t]{0.5\linewidth}
			\centering
			\includegraphics[width=1.5in]{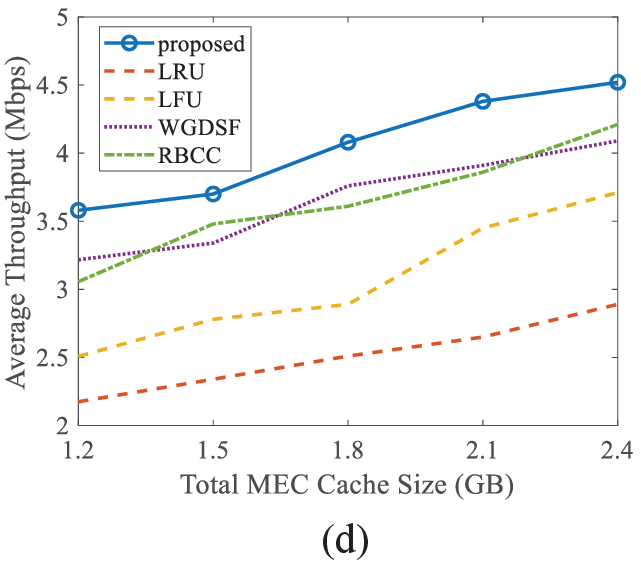}
		\end{minipage}
	}%
	\centering
	\caption{(a)-(c) Throughput of each client for different update algorithms served by MEC server 1, 2 and 3. (d) Average throughput of all clients for different update algorithms vs. different total MEC storage sizes.}
	\label{fig:throughput}
\end{figure}
Fig.\ref{fig:throughput} shows the result of clients' throughput served by different MEC servers and average throughput with different total MEC cache sizes. In Fig.\ref{fig:throughput}.(a), (b) and (c), we present the comparison of client's throughput when the initial cache space of MEC server $1$, $2$ and $3$ is 650MB, 500MB and 550MB respectively. It can be seen that compared with the other four compared algorithms, the clients' throughput of the proposed algorithm is the highest overall, because it comprehensively considers the size of requested segments and the client's current remaining buffer time, and dynamically adjusts segment cache updating strategy to ensure that the client's throughput can meet the smooth playback requirements. Fig.\ref{fig:throughput}.(d) presents the average throughput with the different total MEC cache sizes. On the whole, the average throughput of each algorithm keeps increasing with the increase of the total MEC cache space. In the initial stage, the average throughput of the proposed algorithm is the highest, followed by the WGDSF and RBCC algorithms, and that of the LRU algorithm is the lowest. As the MEC cache space increases, the average throughput of the WGDSF and the RBCC algorithm is relatively close, alternately leading in different MEC cache sizes, but always higher than that of the LFU and LRU algorithm. Because LFU and LRU algorithms simply consider the segment request without considering whether the cached segment representation can match the client's channel state. When the total MEC cache space is $2.4$ GB, the average throughput of each algorithm reaches the maximum, and the average throughput of the proposed algorithm increases by $60.7\%$ , $23.3\%$, $9.8\%$, and $5.9\%$, compared with LRU, LFU, WGDSF and RBCC respectively.

\begin{figure}[htbp]
	\centering
	\subfigure{
		\begin{minipage}[t]{0.5\linewidth}
			\centering
			\includegraphics[width=1.5in]{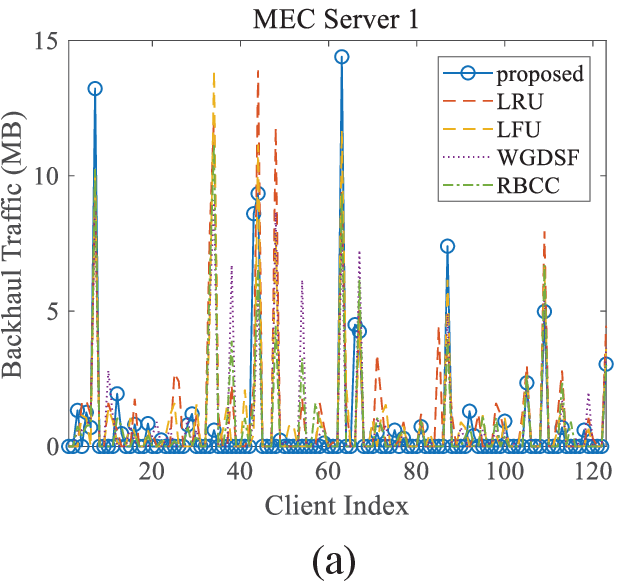}
		\end{minipage}%
	}%
	\subfigure{
		\begin{minipage}[t]{0.5\linewidth}
			\centering
			\includegraphics[width=1.5in]{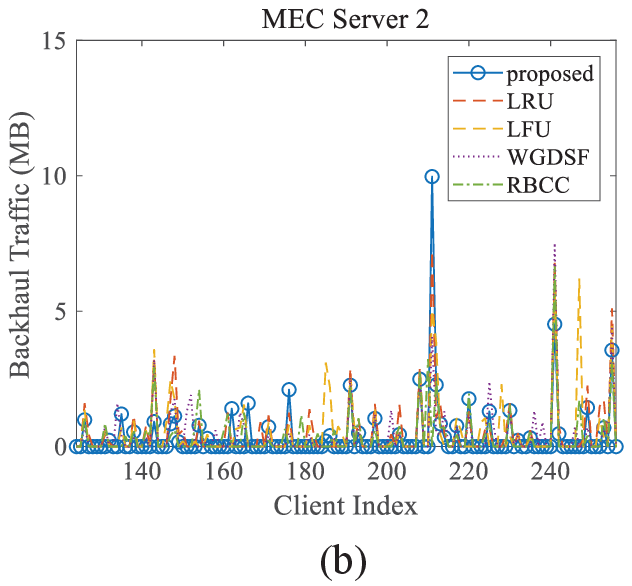}
		\end{minipage}%
	}%

	\subfigure{
		\begin{minipage}[t]{0.5\linewidth}
			\centering
			\includegraphics[width=1.5in]{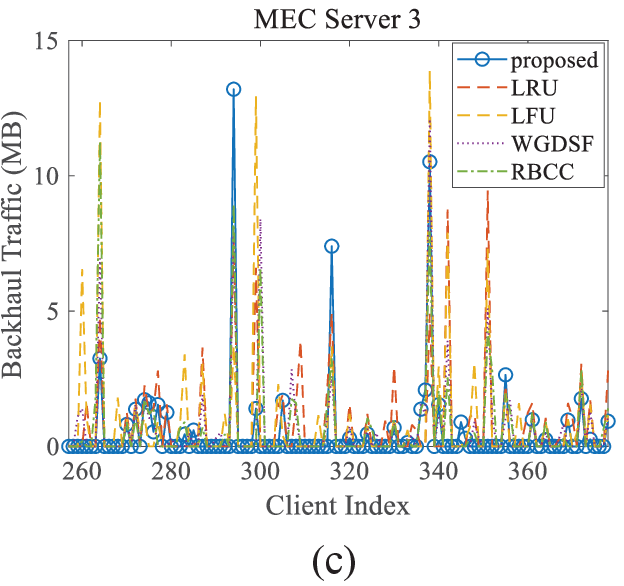}
		\end{minipage}
	}%
	\subfigure{
		\begin{minipage}[t]{0.5\linewidth}
			\centering
			\includegraphics[width=1.5in]{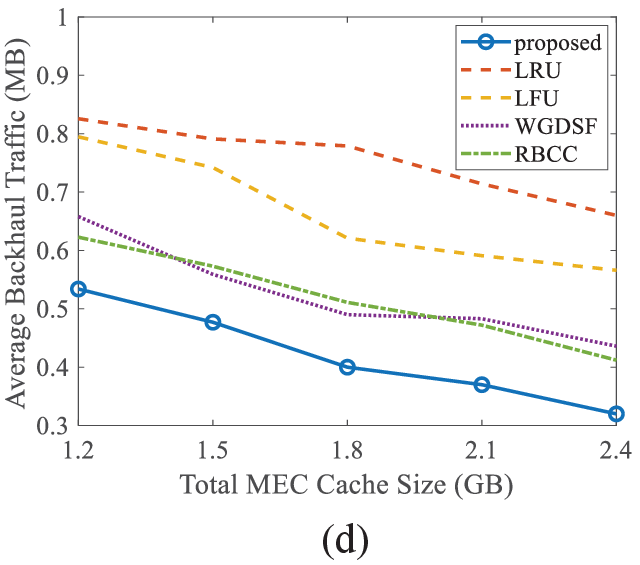}
		\end{minipage}
	}%
	\centering
	\caption{(a)-(c) Backhaul traffic of each client for different update algorithms served by MEC server 1, 2 and 3. (d) Average backhaul traffic of all clients for different update algorithms vs. different total MEC storage sizes.}
	\label{fig:backhaul}
\end{figure}

In Fig.\ref{fig:backhaul}.(a), (b) and (c), we present the comparison of backhaul traffic when the initial cache of MEC server $1$, $2$ and $3$ is 650MB, 500MB and 550MB respectively. On the whole, compared with the other algorithms, although some clients' backhaul traffic of the proposed algorithm is higher, we can still ensure that the backhaul traffic of most clients is lower. Since the proposed algorithm considers the benefits of caching or deleting a segment for the whole in each cache update period, it can cache more suitable segment of the representation for most of the clients. Fig.\ref{fig:backhaul}.(d) presents the average backhaul traffic with different total MEC cache sizes. With the increase of MEC cache space, the average backhaul traffic of each algorithm reduces gradually, and that of proposed algorithm always preserves the lowest. When the total MEC cache space is $2.4$ GB, the average backhaul of each algorithm reaches the minimum level, and the average backhaul of the proposed algorithm decreases by $35.3\%$ , $32.8\%$, $18.8\%$, and $14.2\%$, compared with LRU, LFU, WGDSF and RBCC respectively.

\begin{figure}[htbp]
	\centering
	\subfigure{
		\begin{minipage}[t]{0.5\linewidth}
			\centering
			\includegraphics[width=1.5in]{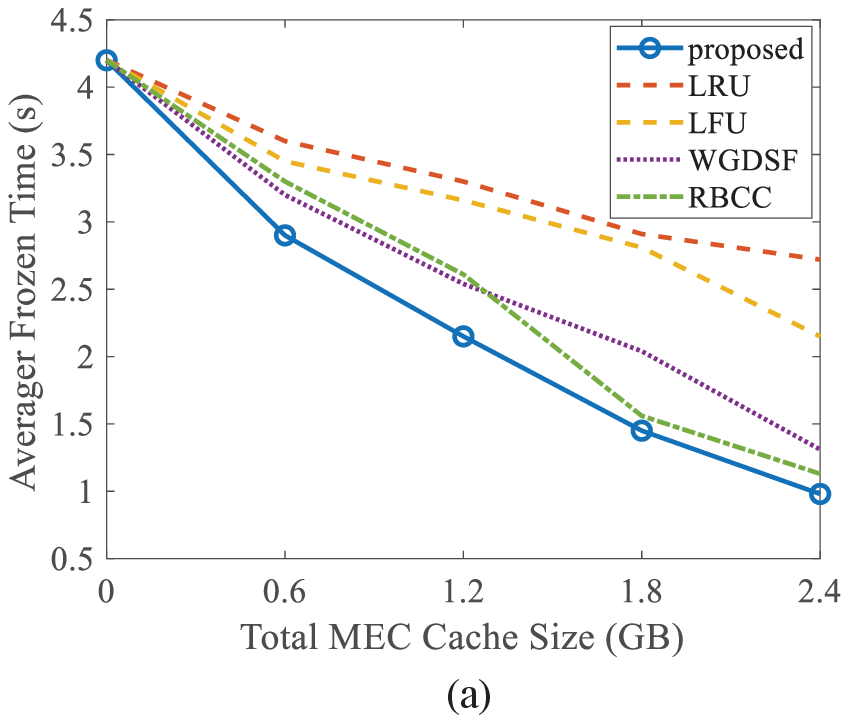}
		\end{minipage}%
	}%
	\subfigure{
		\begin{minipage}[t]{0.5\linewidth}
			\centering
			\includegraphics[width=1.5in]{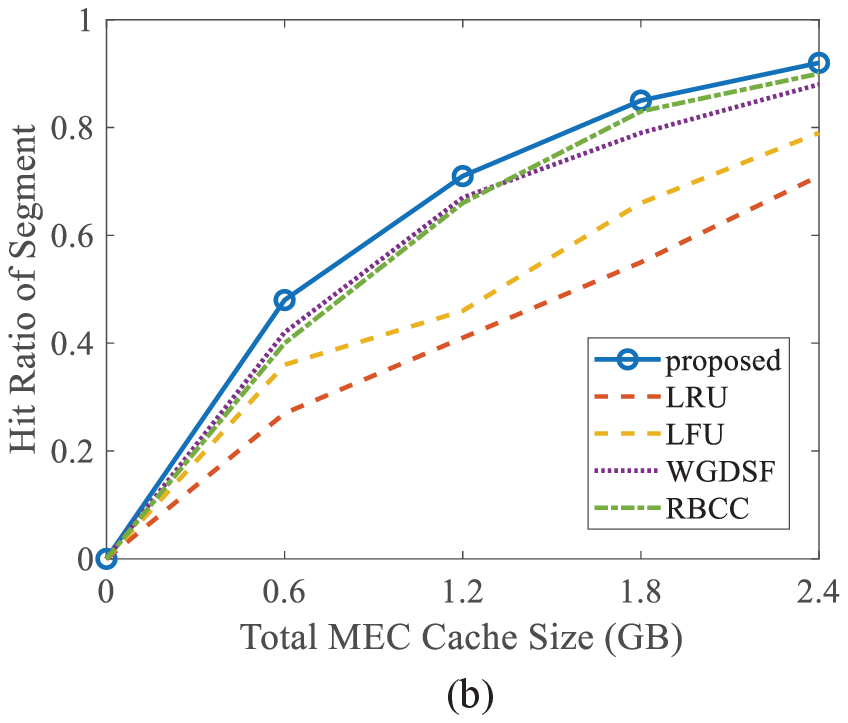}
		\end{minipage}%
	}%
	\centering
	\caption{(a)Average frozen time of all clients for different update algorithms vs. different total MEC cache sizes. (b) Hit ratio of segment of all clients for different update algorithms vs. different total MEC cache sizes.}
	\label{fig:final}
\end{figure}

Fig.\ref{fig:final}.(a) and (b) present the average frozen time and hit ratio of segment with the different total MEC cache sizes, respectively. When the MEC cache is disabled, the average frozen time is 4.2s. This is because the clients have to request the segment from the cloud server through the eNodeB, which will results in large transmission delay due to the long distance. With the increase of the total MEC cache size, the average frozen time of each algorithm reduces and the hit ratio of segment increases. This is because the increase of MEC cache size can provide more space to cache various representations of the segments. From the perspective of the performance of the average frozen time and hit ratio of segment, our proposed algorithm is superior to other compared algorithms.

\section{Conclusion}

In this paper, we propose a more refined video segment based cache update algorithm to improve the QoE of clients and reduce the burden of backhaul in multi-MEC servers, by comprehensively considering the segment popularity, client's transmission state and playback status. First, we divide the MEC cache into different parts based on the content characteristic and popularity of segment. In particular, the size of different cache of the MEC server can be converted to each other, thereby fully utilizing the MEC cache.
And then define the delete priority of segment based on requested number and the relationship between segment rate and client's transmission capacity. Next, we obtain the client's cache priority based on playback status. Finally, we formulate the mathematical model with the objective to maximize the sum of the clients' utility function. The brand and branch method is utilized to obtain the optimal solution to the optimization problem. Simulation results show that the proposed algorithm can achieve a better performance on system throughput, backhaul traffic, frozen time as well as hit ratio of segment compared with other algorithms.

\end{document}